\documentclass[preprint,aps,prb,showpacs,superscriptaddress]{revtex4}
\usepackage{graphicx}
\usepackage{amssymb}
\usepackage{bm}
\begin{document}
\title{Dimensional Crossover of Dilute Neon inside Infinitely Long Single-Walled Carbon Nanotubes Viewed from Specific Heats}
\author{Zhan-chun \surname{Tu}}
\email[Email address:] {tzc@itp.ac.cn} \affiliation{Institute of
Theoretical Physics,
 The Chinese Academy of Sciences,
 P.O.Box 2735 Beijing 100080, China}
  \affiliation{Graduate School,
 The Chinese Academy of Sciences, Beijing, China}
\author{Zhong-can \surname{Ou-Yang}}
\affiliation{Institute of Theoretical Physics,
 The Chinese Academy of Sciences,
 P.O.Box 2735 Beijing 100080, China}
\affiliation{Center for Advanced Study,
 Tsinghua University, Beijing 100084, China}
\begin{abstract}
A simple formula for coordinates of carbon atoms in a unit cell
of a single-walled nanotube (SWNT) is presented and the potential
of neon (Ne) inside an infinitely long SWNT is analytically
derived under the assumption of pair-wise Lennard-Jones potential
between Ne and carbon atoms. Specific heats of dilute Ne inside
infinitely long (5, 5), (10, 10), (15, 15) and (20, 20) SWNT's are
calculated at different temperatures. It is found that Ne inside
four kinds of nanotubes exhibits 3-dimensional (3D) gas behavior
at high temperature but different behaviors at low temperature: Ne
inside (5, 5) nanotube behaves as 1D gas but inside (10, 10), (15,
15), and (20, 20) nanotubes behaves as 2D gas. Furthermore, at
ultra low temperature, Ne inside (5, 5) nanotube still displays 1D
behavior but inside (10, 10), (15, 15), and (20, 20) nanotubes
behaves as lattice gas.
\end{abstract}
 \maketitle
Since the discovery of carbon nanotubes,\cite{s} the
peculiar electronic and mechanical properties of these
structures have attracted much attention. \cite{jw,bi,x}
Experiments have also revealed that carbon nanotubes can be
used to store hydrogen \cite{dillon} and other gases.
\cite{teizer} Many physicists expect gases in nanotubes or
nanotube bundles to display 1-dimensional (1D) behavior as
a consequence of the remarkable aspect ratio of the length
of tubes to their radius. Cole and his co-workers
\cite{cole} and other researchers \cite{carraro} have
theoretically studied properties of gases in nanotubes or
nanotube bundles. One of the most fantastic properties they
found is specific heat of dilute gas inside single-walled
carbon nanotubes (SWNT's) as a functions of temperature:
With temperature increasing it shows the thermal behavior
changing from 1D to 2D, to 3D. However, there is still a
question: Do 1D and 2D behaviors always exist for dilute
gas inside SWNT's if the temperature is low enough?

A series of skillful experiments to measure specific heat
of gas in carbon nanotube bundles were done by Lasjaunias
{\it et al}.\cite{Lasjaunias} In their experiments, He
atoms are adsorbed within the interstitials or external
grooves and surfaces of the bundles, and exhibit 1D
behavior at low temperature. Their experiments suggest that
the specific heat can reflect the dimensional information
well.

Here we firstly present a simple formula for coordinates of
carbon atoms in a unit cell of a SWNT and derive an
analytical expression of the potential of neon (Ne) inside
an infinitely long SWNT under the assumption of pair-wise
Lennard-Jones potential between Ne and carbon atoms. And
then, we calculate specific heats of dilute Ne inside (5,
5), (10, 10), (15, 15) and (20, 20) SWNT's and find that Ne
inside four kinds of nanotubes exhibits 3D gas behavior at
high temperature but different behaviors at low
temperature: Ne inside (5, 5) nanotube behaves as 1D gas
but inside (10, 10), (15, 15), and (20, 20) nanotubes
behaves as 2D gas. Furthermore, at ultra low temperature,
Ne inside (5, 5) nanotube still displays 1D behavior but
inside (10, 10), (15, 15), and (20, 20) nanotubes behaves
as lattice gas. Ne inside (5, 5) nanotube does not display
2D behavior and inside (10, 10), (15, 15), and (20, 20)
nanotubes does not display 1D behavior. Thus Ne inside
nanotubes does not always display 1D behavior in spite of
the remarkable aspect ratio of the length of tubes to their
radius. Besides, Ne inside a definite nanotube may display
neither 1D nor 2D behaviors at different temperatures.

\begin{figure}[htp!]
\includegraphics[width=7cm]{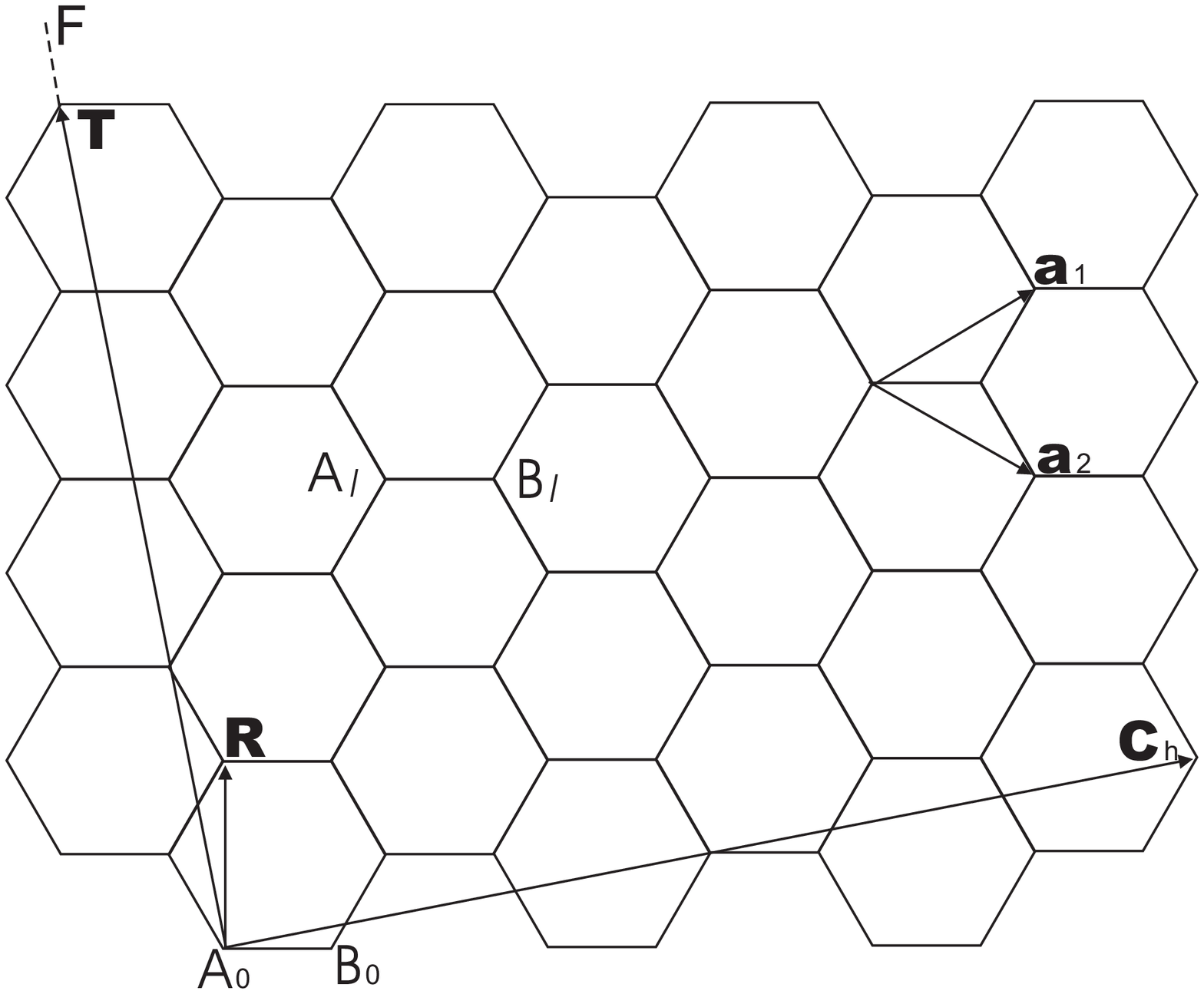}
\caption{\label{fig1}The unrolled honeycomb lattice of a SWNT. By
rolling up the sheet along the chiral vector ${\bf C}_h$, that is,
such that the point $A_0$ coincides with the point corresponding
to vector ${\bf C}_h$, a nanotube is formed. The vectors ${\bf
a}_{1}$ and ${\bf a}_{2}$ are the real space unit vectors of the
hexagonal lattice. The translational vector ${\bf T}$ is
perpendicular to ${\bf C}_h$ and runs in the direction of the tube
axis. The vector ${\bf R}$ is the symmetry vector. $A_0$, $B_0$
and $A_l, B_l (l=1,2,\cdots,N)$ are used to denote the sites of
carbon atoms.}
\end{figure}

A SWNT without two caps can be constructed by wrapping up a single
sheet of graphite such that two equivalent sites of hexagonal
lattice coincide. \cite{r1} To describe the SWNT, some
characteristic vectors require introducing. As shown in
Fig.\ref{fig1}, the chiral vector ${\bf C}_{h}$, which defines the
relative location of two sites, is specified by a pair of integers
$(n_1, n_2)$ which is called the index of the SWNT and relates
${\bf C}_{h}$ to two unit vectors ${\bf a}_{1}$ and ${\bf a}_{2}$
of graphite (${\bf C}_{h}=n_1{\bf a}_{1}+n_2{\bf a}_{2}$). The
translational vector ${\bf T}$ corresponds to the first lattice
point of 2D graphitic sheet through which the line normal to the
chiral vector ${\bf C}_{h}$ passes. The unit cell of the SWNT is
the rectangle defined by vectors ${\bf C}_{h}$ and ${\bf T}$,
while vectors ${\bf a}_{1}$ and ${\bf a}_{2}$ define the area of
the unit cell of 2D graphite. The number $N$ of hexagons per unit
cell of SWNT is obtained as a function of $n_1$ and $n_2$ as
$N=2(n_1^2+n_2^2+n_1n_2)/d_R$, where $d_R$ is the greatest common
divisor of ($2n_2+n_1$) and ($2n_1+n_2$). There are $2N$ carbon
atoms in each unit cell of SWNT because every hexagon contains two
atoms. To denote the $2N$ atoms, we use a symmetry vector ${\bf
R}$ to generate coordinates of carbon atoms in the nanotube and
define it as the site vector having the smallest component in the
direction of ${\bf C}_h$. From a geometric standpoint, vector
${\bf R}$ consists of a rotation around the nanotube axis by an
angle $\Psi=2\pi/N$ combined with a translation $\tau$ in the
direction of ${\bf T}$; therefore, ${\bf R}$ can be denoted by
${\bf R}=(\Psi|\tau)$. Using the symmetry vector ${\bf R}$, we can
divide the $2N$ carbon atoms in the unit cell of SWNT into two
groups: one includes $N$ atoms whose site vectors satisfy
\begin{equation}\label{sitea}
{\bf A}_l=l{\bf R}-[l{\bf R}\cdot{\bf T}/{\bf T}^2]{\bf T} \quad
(l=0,1,2,\cdots,N-1),\end{equation} another includes the remainder
$N$ atoms whose site vectors satisfy
\begin{eqnarray}\label{siteb} {\bf B}_l&=&l{\bf R}+{\bf
B}_0-[(l{\bf R}+{\bf B}_0)\cdot{\bf T}/{\bf T}^2]{\bf T}\nonumber
\\ &-&[(l{\bf R}+{\bf B}_0)\cdot{\bf C}_h/{\bf C}_h^2]{\bf C}_h
\quad (l=0,1,2,\cdots,N-1),\end{eqnarray}
 where {\bf B}$_0$
represents one of the nearest neighbor atoms to {\bf A}$_0$. In
and only in above two equations, $[\cdots]$ denotes the Gaussian
function, e.g., $[5.3]=5$.

\begin{figure}[htp!]
\includegraphics[width=7cm]{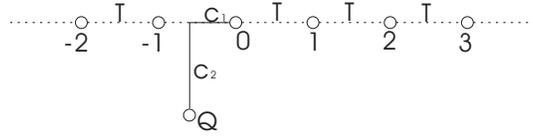}
\caption{\label{fig2}An infinite atom chain and an atom Q out of
the chain. Many atoms distribute regularly in a line form the
infinite atom chain. The interval between neighbor atoms in the
chain is $T$, and the site of atom Q relative to atom 0 can be
represented by numbers $c_1$ and $c_2$.}
\end{figure}

To obtain the potential of Ne inside the nanotube, we
firstly consider another simple system shown in
Fig.~\ref{fig2}: Many carbon atoms distributed regularly in
a line form an infinite atom chain and a Ne atom Q is out
of the chain. The interval between neighbor atoms in the
chain is $T$, and the site of atom Q relative to atom 0 can
be represented by two numbers $c_1$ and $c_2$. We take the
Lennard-Jones potential
$U(R_j)=4\epsilon[(\sigma/R_j)^{12}-(\sigma/R_j)^6]$
between atom Q and atom $j$ in the chain, where $R_j$ is
the distance between Q and atom $j$, and
$\epsilon=\sqrt{\epsilon_c\epsilon_{ne}}$,
$\sigma=(\sigma_c+\sigma_{ne})/2$ with $\epsilon_{ne}=35.6$
K, $\sigma_{ne}=2.75$ \AA,\ \ $\epsilon_c=28$ K and
$\sigma_c=3.4$ \AA. \cite{cole,hir} We express the
potential between atom Q and the chain as
\begin{equation} \label{upc}
U_{QC}=4\epsilon[\sigma^{12}U_6(c_1,c_2)-\sigma^6U_3(c_1,c_2)]
,\end{equation}where
$U_k(c_1,c_2)=\sum\limits_{n=-\infty}^{\infty}\frac{1}{[(c_1+nT)^2+c_2^2]^k}\quad
(k=1,2,\cdots)$ which can be calculated through the following
recursion: \cite{anran}
\begin{equation}\label{uk}
\begin{array}{l} U_1(c_1,c_2)=\frac{\pi
\sinh (2\pi c_2/T)}{c_2T[\cosh (2\pi c_2/T)-\cos (2\pi c_1/T)]},\\
U_{k+1}(c_1,c_2)=-1/(2kc_2) \partial U_k/\partial c_2. \end{array}
\end{equation}

\begin{figure}[htp!]
\includegraphics[width=7cm]{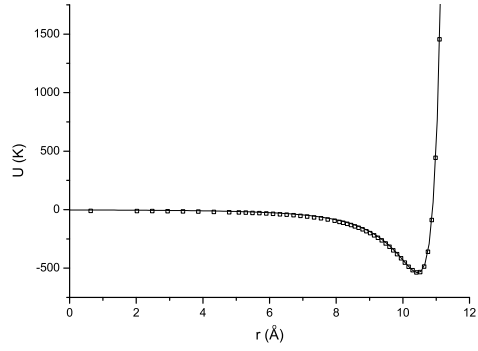}
\caption{\label{fig3} The potentials inside the $(20,20)$ nanotube
calculated from Eq.(\ref{potential}) (squires) and the fit curve
(solid curve).}
\end{figure}

The $(20, 20)$ tube, for example, with infinite length can be
regarded as $2N=80$ chains. Thus the potential of any point Q inside
the tube can be calculated as
\begin{equation}\label{potential}U(r,\theta,z)=\sum_{i=1}^{2N}U_{QC},\end{equation}
where $(r,\theta,z)$ is coordinates of Q in the cylindrical
coordinate system whose $z$-axis is the tube axis, $r$ the
distance between Q and $z$-axis, and $\theta$ the angle rotating
around $z$-axis from an axis which is vertical to $z$-axis and
passes through atom $A_0$ on the tube to the plane that contains Q
and $z$-axis. As an approximation, we neglect the potential
varying with $z$ and $\theta$ because we find that it is much
smaller than the potential varying with $r$ through calculations,
and fit the potential with
$U(r)=4\varepsilon[(\frac{\tilde{\sigma}}{\rho-r})^{10}-(\frac{\tilde{\sigma}}{\rho-r})^{4}]$,
where $\rho=13.56$ \AA\ \ is the radius of the tube,
$\varepsilon=390$ K, and $\tilde{\sigma}=2.63$ \AA \ \ (see also
Fig.\ref{fig3}). Moreover, we simplify it as
\begin{equation}\label{ur}
U(r)=\left
\{\begin{array}{l}4\varepsilon[(\frac{\tilde{\sigma}}{\rho-r})^{10}-(\frac{\tilde{\sigma}}{\rho-r})^{4}]
,\quad r<\rho-\tilde{\sigma},\\
\infty,\quad r>\rho-\tilde{\sigma}.\end{array}\right.
\end{equation}

Because we consider the dilute Ne, we can neglect the interaction
between Ne atoms and write the single particle Schr\"{o}dinger
equation \cite{landau1} as $H\psi=E\psi$, where
$H=-\frac{\hbar^2}{2\mu}\nabla^2+U(r)$ and $\psi=\phi
e^{i(m\theta+\kappa z)}$. From them we arrive at
\begin{equation}\label{schr2}
\label{schr} \begin{array}{l}E=\frac{\hbar^2\kappa^2}{2\mu}+E_m\quad (\kappa\in \mathbb{R}, m=0,\pm 1,\pm 2,\cdots),\\
H(r)\phi=E_m\phi,\\
H(r)=-\frac{\hbar^2}{2\mu}(\frac{d^2}{dr^2}+\frac{1}{r}\frac{d}{dr}-\frac{m^2}{r^2})+U(r).\end{array}
\end{equation}
Setting $r=(\rho-\tilde{\sigma})\xi$,
$\varepsilon_0=\frac{\hbar^2}{2\mu\rho^2}$ and
$\eta=\tilde{\sigma}/\rho$, Eqs.(\ref{ur}) and (\ref{schr2}) are
transformed into
\begin{equation}\label{uxi}
u(\xi)=\left
\{\begin{array}{l}4\varepsilon[(\frac{\eta}{1-(1-\eta)\xi})^{10}-(\frac{\eta}{1-(1-\eta)\xi})^{4}]
,\quad \xi<1,\\
\infty,\quad \xi> 1.\end{array}\right.
\end{equation}
and
\begin{equation}\label{schr3}
\label{schr} \begin{array}{l}\mathcal{H}\varphi(\xi)=E_m\varphi(\xi),\\
\mathcal{H}=-\frac{\varepsilon_0}{(1-\eta)^2}(\frac{d^2}{d\xi^2}+\frac{1}{\xi}\frac{d}{d\xi}-\frac{m^2}{\xi^2})+u(\xi).\end{array}
\end{equation}
If we let $|\varphi\rangle=\sum_na_n|\chi_n\rangle$, we will
obtain the secular equation
\begin{equation}\label{scu1}
det(\mathcal{H}_{jn}-E_m\mathcal{S}_{jn})=0,
\end{equation}
where $\mathcal{H}_{jn}=\int_0^1\chi_j\mathcal{H}(\xi)\chi_n\xi
d\xi$, and $\mathcal{S}_{jn}=\int_0^1\chi_j\chi_n\xi d\xi$. If we
let $\chi_n=J_{|m|}(\nu_n\xi)$, where $J_{|m|}(\xi)$ is the m-th
order Bessel function of the first class and $\nu_n$ is the n-th
zero point of Bessel function,\cite{wang} we can calculated
$E_{mn}, (m=0,\pm 1,\pm 2,\cdots; n=1,2,3,\dots)$ from
Eq.(\ref{scu1}).

If there are $\mathcal{N}$ Ne atoms inside the tube, we have the
free energy $\mathcal{F}=-\mathcal{NT}\ln\mathcal{Z}$, where
$\mathcal{T}$ is the temperature,
$\mathcal{Z}=\sum_{mn}e^{-E_{mn}/\mathcal{T}}\int_{-\infty}^{\infty}
e^{-\frac{\hbar^2\kappa^2}{2\mu\mathcal{T}}}d\kappa$ and the
Boltzmann factor is set to 1.\cite{landau2} We can easily obtain
the specific heat per atom is
\begin{equation}
c_v=-\frac{\mathcal{T}\partial^2\mathcal{F}}{\mathcal{N}\partial\mathcal{T}^2}=\frac{1}{2}+\frac{\langle
E^2\rangle-\langle E\rangle^2}{\mathcal{T}^2},
\end{equation}
where $\langle E\rangle=\frac{\sum_{mn}
E_{mn}e^{-E_{mn}/\mathcal{T}}}{\sum_{mn}e^{-E_{mn}/\mathcal{T}}}$
and $\langle E^2\rangle=\frac{\sum_{mn}
E^2_{mn}e^{-E_{mn}/\mathcal{T}}}{\sum_{mn}e^{-E_{mn}/\mathcal{T}}}$.

\begin{figure}[htp!]
\includegraphics[width=7cm]{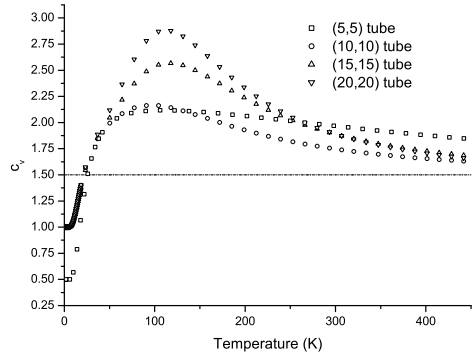}
\caption{\label{fig4} The specific heats per atom $c_v$ of Ne
inside $(5,5)$, $(10,10)$, $(15,15)$ and $(20,20)$ nanotubes under
the approximate potentials varying with the temperature.}
\end{figure}

In Fig.\ref{fig4}, the symbols ``$\triangledown$" reflect $c_v$
varying with temperature $\mathcal{T}$, which implies Ne
inside (20,20) tube behaves as 3D gas at high temperature (specific
heat approaches to 3/2) and 2D gas at low temperature (specific
heat is 1). Therefore, we can naturally assume that all atoms in
the valley of potential $U(r)$ at low temperature, i.e. lie on the
shell $S^*$ with radius $\varrho=\rho[1-(5/2)^{1/6}\eta]$.

Now we consider the thermal property of Ne inside the nanotube at
low temperature more carefully. From
Eqs.(\ref{upc})-(\ref{potential}) we can easily calculate the
potential $U_s(v,z)$ on $S^*$, where $v=\varrho \theta$. The
Hamiltonian of single particle can be expressed as $
H'=-\frac{\hbar^2}{2\mu}(\frac{\partial^2}{\partial
v^2}+\frac{\partial^2}{\partial z^2})+U_s(v,z).
$

In fact, $U_s$ has periodic structure. If we denote ${\bm
\alpha}_1=(2\pi\varrho/N,\tau)$, ${\bm \alpha}_2=(0,T)$ and ${\bm
\gamma}_l=l_1{\bm \alpha}_1+l_2{\bm \alpha}_2$, we know $U_s({\bf
r}+{\bm \gamma}_l)=U_s({\bf r})$, where $l_1,l_2\in \mathbb{Z}$
and ${\bf r}=(v,z)$. On the one hand, we have the Bloch's theorem:
\cite{kittel}
\begin{equation}\label{plane}
\begin{array}{l}
H'\Phi({\bm \kappa},{\bf r})=E_{\bm \kappa}\Phi({\bm \kappa},{\bf r}),\\
\Phi({\bm \kappa},{\bf
r}+{\bm\gamma}_l)=e^{i{\bm\kappa}\cdot{\bm\gamma}_l}\Phi({\bm\kappa},{\bf
r}),\end{array}
\end{equation}
which suggests that $\Phi({\bm \kappa},{\bf
r})=\sum_ja({\bm\kappa}+{\bf G}_j)e^{i({\bm\kappa}+{\bf G}_j)\cdot
{\bf r}}$, where ${\bf G}_j=j_1{\bm \beta}_1+j_2{\bm \beta}_2$
with $j_1,j_2\in \mathbb{Z}$, ${\bm \beta}_1=(N/\varrho,0)$ and
${\bm \beta}_2=(-\tau N/(T\varrho),2\pi/T)$. From
Eq.(\ref{plane}), we obtain the secular equation
\begin{equation}\label{scu2}
det(\mathcal{H}_{lj}-E_{\bm \kappa}\delta_{lj})=0,
\end{equation}
where $\mathcal{H}_{lj}=\frac{\hbar^2}{2\mu}({\bm\kappa}+{\bf
G}_j)^2\delta_{lj}+\mathcal{U}_{lj}$,
$\mathcal{U}_{lj}=\frac{1}{\Omega_0}\int_{\Omega_0}e^{i({\bf
G}_l-{\bf G}_j)\cdot {\bf r}}U_s({\bf r})d{\bf r}$ and
$\Omega_0=|{\bm \alpha}_1\times{\bm \alpha}_2|$.

On the other hand, Periodic boundary condition along the
circumference of the shell $S^*$ suggests that we just need to
consider the first Brillouin zone which consists of ${\bm
\kappa}=(m_v/\varrho,\kappa_z)$ where $m_v\in\mathbb{Z}, 0\leq
m_v<N$ and $\kappa_z\in\mathbb{R}, 0\leq \kappa_z<2\pi/T$.

From Eq.(\ref{scu2}), we can calculate the energy
$E_{m_v,\kappa_z}$, and then the free energy
$\mathcal{F}=-\mathcal{NT}\ln\mathcal{Z}$,\cite{remark1} where
$\mathcal{Z}=\sum_{m_v}\int_0^{2\pi/T}e^{-E_{m_v,\kappa_z}/\mathcal{T}}d\kappa_z$.
Moreover, the specific heat per atom is \cite{landau2}
\begin{equation}
c_v=-\frac{\mathcal{T}\partial^2\mathcal{F}}{\mathcal{N}\partial\mathcal{T}^2}=\frac{\langle
E'^2\rangle-\langle E'\rangle^2}{\mathcal{T}^2},
\end{equation}
where $\langle
E'\rangle=\sum\limits_{m_v=0}^{N-1}\int_0^{2\pi/T}E_{m_v,\kappa_z}e^{-E_{m_v,\kappa_z}/\mathcal{T}}d\kappa_z$
and $\langle
E'^2\rangle=\sum\limits_{m_v=0}^{N-1}\int_0^{2\pi/T}E^2_{m_v,\kappa_z}e^{-E_{m_v,\kappa_z}/\mathcal{T}}d\kappa_z$.

\begin{figure}[htp!]
\includegraphics[width=7cm]{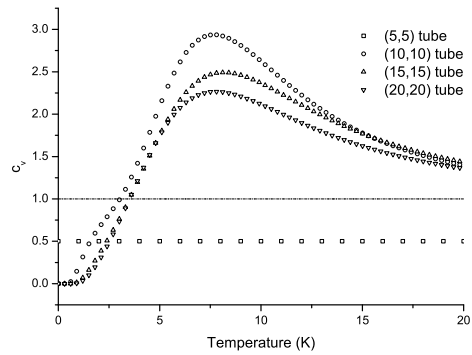}
\caption{\label{fig5}The specific heat per atom $c_v$ of Ne inside
$(5,5)$, $(10,10)$, $(15,15)$ and $(20,20)$ nanotubes at low
temperatures.}
\end{figure}

In Fig.\ref{fig5}, the symbols ``$\triangledown$" reflect $c_v$
varying with the temperature $\mathcal{T}$, which implies Ne atoms
inside (20,20) tube behave as the lattice gas \cite{cole2} at
ultra low temperature (specific heat is 0) and 2D gas at low
temperature (specific heat approaches 1). There is no 1D gas
inside (20,20) tube, which is quite different from our usual
notion.

Adopting the similar method, we calculate $c_v$ of Ne
inside $(5,5)$, $(10,10)$ and $(15,15)$ nanotubes and show
them in Fig.\ref{fig4} and Fig.\ref{fig5}. When we obtain
Fig.\ref{fig4}, we have adopted the approximate potentials
as
$U(r)=4\varepsilon[(\frac{\tilde{\sigma}}{\rho-r})^{p_1}-(\frac{\tilde{\sigma}}{\rho-r})^{p_2}]$.
The parameters $\varepsilon$, $\tilde{\sigma}$, $\rho$,
$p_1$, and $p_2$ are shown in Table \ref{tab1}. These two
figures imply that Ne inside four kinds of nanotubes
exhibits 3D gas behavior at high temperature but different
behaviors at low temperature: Ne inside (5, 5) nanotube
behaves as 1D gas (specific heat is 1/2) but inside (10,
10), (15, 15), and (20, 20) nanotubes behaves as 2D gas.
Furthermore, at ultra low temperature, Ne inside (5, 5)
nanotube still displays 1D behavior but inside (10, 10),
(15, 15), and (20, 20) nanotubes behaves as lattice gas. Ne
inside (5, 5) nanotube does not display 2D behavior and
inside (10, 10), (15, 15), and (20, 20) nanotubes does not
display 1D behavior. Thus Ne inside nanotubes does not
always display 1D behavior in spite of the remarkable
aspect ratio of the length of tubes to their radius.
Besides, Ne inside a definite nanotube may not display both
1D and 2D behaviors at different temperatures.

Our calculations support the view of dimensional crossover
of gas inside SWNT's put forward by Cole {\it et.al.}
\cite{cole} but there is a significant difference: Cole
{\it et.al.} obtain the dimensionality varying from one to
two to three with the temperature increasing, but we obtain
the dimensionality varying either from one directly to
three or from two to three. The reason is that the
potential of gas inside SWNT's selected by Cole {\it
et.al.} departs father from the real case than that done by
us.

If the potential on shell $S^*$ is uniform, one can
intuitively expect the dimensionality to vary from one to
two to three as the axial, azimuthal and radial degrees of
freedom become excited as temperature increases. But the
potential on this shell has more exquisite structure (the
periodic structure) for large diameter nanotubes, such as
(10,10), (15,15) and (20,20) tubes, which results in the
approximate same axial and azimuthal excited energy.
Therefore Ne in large diameter nanotubes exhibits lattice
gas (0D), 2D and 3D behavior as temperature increases. On
the contrary, the shell converges to the tube axis for
small nanotube, e.g. (5,5) tube, which results in the
dimensionality varying from one direct to three as
temperature increases.

The authors acknowledge the useful discussions with Prof.
H. W. Peng, Dr. M. Li, F. Liu, Y. Zhang, L. R. Dai, and J.
J. Zhou. Furthermore, we thank Dr. H. J. Zhou because he
help us read the English.

\newpage

\begin {table}[htp!]
\caption{\label{tab1}The parameters in the approximate potentials
for different nanotubes.}
\begin {ruledtabular}
\begin {tabular} {cccccc}
(n,m)&$\varepsilon$ (K)& $\tilde{\sigma}$ (\AA)& $\rho$ (\AA)& $p_1$& $p_2$ \\
\hline
(5,5)&520 & 2.63& 3.39&10 &1 \\
(10,10)& 360 &2.63 &6.78 &10 &3 \\
(15,15)&418 &2.64 & 10.17& 10 & 4\\
\end{tabular}
\end {ruledtabular}
\end{table}

\begin{thebibliography}{}
\bibitem{s}S. Iijima, Nature (London) {\bf 354}, 56~(1991).
\bibitem{jw}N. Hamada, S. I. Sawada, and A. Oshiyama, Phys. Rev. Lett. {\bf 68}, 1579~(1992); R. Saito, M. Fujita, G. Dresselhaus, and M. S. Dresselhaus, Appl. Phys. Lett. {\bf 60},
2204~(1992); J. W. Mintmire, B. I. Dunlap, and C. T. White, Phys.
Rev. Lett. {\bf 68}, 631~(1992).
\bibitem{bi}B. I. Yakobson and P. Avouris, {\it Mechanical Properties
of Carbon Nanotubes, in Carbon Nanotubes}, edited by M. S.
Dresselhaus and P. Avouris (Springer-Verlag, Berlin, 2001), pp.
287-327.
\bibitem{x}Z. C. Ou-Yang, Z. B. Su, and C. L. Wang, Phys. Rev. Lett. {\bf 78}, 4055~(1997); X. Zhou, J. J. Zhou, and Z. C. Ou-Yang, Phys. Rev. B {\bf 62},
13692~(2000); Z. C. Tu and Z. C. Ou-Yang, Phys. Rev. B {\bf 65},
233407 (2002).
\bibitem{dillon}A. C. Dillon, K. M. Jones, T. A. Bekkedahl, C. H. Kiang, D. S. Bethune, and M. J. Heben, Nature {\bf 386}, 377
(1997); C. Liu, Y. Y. Fan, M. Liu, H. T. Cong, H. M. Cheng, and M.
S. Dresselhaus, Science {\bf 286}, 1127 (1999).
\bibitem{teizer}W. Teizer, R. B. Hallock, E. Dujardin, and T. W.
Ebbesen, Phys. Rev. Lett. {\bf 82}, 5305 (1999); {\bf 84}, 1844
(2000); A. Kuznetsova, J. T. Yates, Jr., J. Liu, and R. E.
Smalley, J. Chem. Phys. {\bf 112}, 9590 (2000); S. Talapatra, A.
Zambano, S. E. Weber, and A. D. Migone, Phys. Rev. Lett. {\bf 85},
138 (2000).
\bibitem{cole}M. Calbi, M. W. Cole, S. Gatica, M. J. Bojan and G.
Stan, Rev. Mod. Phys. {\bf 73}, 857 (2001); M. Calbi and M. W.
Cole, Phys. Rev. B {\bf 66}, 115413 (2002); S. Gatica, M. J.
Bojan, G. Stan and M. W. Cole, J. Chem. Phys. {\bf 114}, (2001);
M. Calbi, F. Toigo and M. W. Cole, Phys. Rev. Lett. {\bf 86}, 5062
(2001); G. Stan, M. W. Cole, J. Hartman, V. H. Crespi and S.
Gatica, Phys. Rev. B {\bf 61}, 7288 (2000); G. Stan, M. J. Bojan,
S. Curtarola, S. Gatica and M. W. Cole, Phys. Rev. B {\bf 62},
2173 (2000); G. Stan, S. Gatica, M. Boninsegni, S. Curtarola and
M. W. Cole, Am. J. Phys. {\bf 67}, 1170 (1999); G. Stan and M. W.
Cole, J. Low Temp. Phys. {\bf 110}, 539 (1998); G. Stan, V. H.
Crespi, M. W. Cole and M. Boninsegni, J. Low Temp. Phys. {\bf
113}, 447 (1998).
\bibitem{carraro}C. Carraro, Phys. Rev. B {\bf 61}, R16351 (2000);
A. J. Zambano, S. Talapatra, and A. D. Migone, Phys. Rev. B {\bf
64}, 75415 (2001); M. C. Gordillo, J. Boronat, and J. Casulleras,
Phys. Rev. Lett. {\bf 85}, 2348 (2000).
\bibitem{Lasjaunias}J. C. Lasjaunias, K. Biljakovi\'c, J. L. Sauvajol, and P. Monceau, Phys. Rev. Lett. {\bf 91}, 25901 (2003);
J. C. Lasjaunias, K. Biljakovic\', Z. Benes, J. E. Fischer, and P. Monceau, Phys. Rev. B {\bf 65}, 113409 (2002).
\bibitem{r1}R. Saito, M. S. Dresselhaus, and G. Dresselhaus, {\it Physical Properties of Carbon Nanotubes} (Imperial College Press, London, 1998).
\bibitem{hir}J. O. Hirschfelder, C. F. Curtiss, and R. B. Bird,
{\it Molecular Theory of Gases and Liquids} (John Wiley \& Sons,
Inc., New York, 1954).
\bibitem{anran}R. An (private communication); H. H. Peng and X. S.
Xu, {\it The Fundamentals of Theoretical Physics} (Peking
University Press, Beijing, 1998).
\bibitem{landau1}L. D. Landau and E. M. Lifshitz, {\it Quantum
Mechanics} (Pergamon, Oxford, 1986).
\bibitem{wang}Z. X. Wang and D. R. Guo, {\it Introduction to
Special Function} (Peking University Press, Beijing, 2000).
\bibitem{landau2}L. D. Landau and E. M. Lifshitz, {\it Statistical
Physics} (Pergamon, Oxford, 1986); L. E. Reichl, {\it A Modern
Course in Statistical Physics} (John Wiley \& Sons, Inc., New
York, 1998).
\bibitem{kittel}C. Kittel, {\it Introduction to Solid State
Physics} (John Wiley \& Sons, Inc., New York, 1996).
\bibitem{remark1}In fact, here we just consider the energies in
the first Brillouin zone. If considering $E_{\bm \kappa}=E_{-\bm
\kappa}$ and $E_{\bm \kappa}=E_{{\bm\kappa}+{\bf G}_j}$, we need
to multiple a constant before the partition function $\mathcal{Z}$
in the expression of the free energy.
\bibitem{cole2}M. M. Calbi, S. M. Gatica, M. J. Bojan, and M. W.
Cole, cond-mat/0209220.
\end{thebibliography}
\end{document}